\documentclass{aa}
\usepackage{natbib}
\usepackage{epsfig}
\usepackage{graphicx}
\def\ut#1{\mathop{\vtop{\ialign{##\crcr
     $\hfil\displaystyle{#1}\hfil$\crcr\noalign
     {\kern1pt\nointerlineskip}\hbox{$\hfil\sim\hfil$}\crcr
     \noalign{\kern1pt}}}}}

\def\undersymbol#1#2{\mathop{\vtop{\ialign{##\crcr
     $\hfil\displaystyle{#2}\hfil$\crcr\noalign
     {\kern1pt\nointerlineskip}\hbox{$\hfil#1\hfil$}\crcr
     \noalign{\kern1pt}}}}}
\def\arcsec{^{\prime\prime}}
\def\arcmin{^{\prime}}
\def\degr{^0}

\begin{document}

 \title{The globular cluster NGC 6388: $XMM$-Newton and $Chandra$ observations}

      \author{A.A. Nucita\inst{1}, F. De Paolis\inst{2}, G. Ingrosso\inst{2}, S. Carpano\inst{1}, \and M. Guainazzi\inst{1}}
        \institute{XMM-Newton Science Operations Centre, ESAC, ESA, PO Box 78, 28691 Villanueva de la $\rm Ca\tilde{n}ada$, Madrid, Spain \and
          Dipartimento di Fisica, Universit\`a del Salento, and {\it INFN}, Sezione di Lecce, CP 193, I-73100 Lecce, Italy}

   \offprints{A. A. Nucita}

   \date{Submitted: XXX; Accepted: XXX}

{
  \abstract
   {By studying the optical brightness surface density of the globular cluster NGC 6388, it has been recently proposed that it harbors a central intermediate-mass black hole with mass $\simeq 5.7\times 10^3$ M$_{\odot}$.}
  {We expect that the compact object in the center of NGC 6388 emits radiation in the $X$-ray band as a consequence of the  accretion from the surrounding matter. We searched for $XMM$-Newton and $Chandra$ observations towards NGC 6388 to test this hypothesis.}
   {We determine both the hardness ratios and luminosity with a minimum set of assumptions for each of the identified field sources.}
   {The $Chandra$ satellite disentangles several point-like $X$-ray sources, probably low mass $X$-ray binaries, well within the core radius of the globular cluster.  However, three of them, coinciding with the cluster center of gravity, remain unresolved. Their total luminosity is $L_X^{Obs}\simeq 2.7\times 10^{33}$ erg s$^{-1}$. If one of these sources is the $X$-ray counterpart of the intermediate-mass black hole in NGC 6388, the corresponding upper limit on the accretion efficiency, with respect to the Eddington luminosity, is $3\times 10^{-9}$. This measurement could be tightened  if moderately deep radio observations of the field were performed.}
   {}
}

   \authorrunning{Nucita et al.}
   \titlerunning{$XMM$-Newton and $Chandra$ observation of NGC 6388}
   \maketitle
%


\section{Introduction}

Over the last few years, several pieces of evidence have been accumulated pointing to the
presence of an intermediate-mass black hole (hereafter IMBH) with mass
$\simeq 10^3$ M$_{\odot}$ in a globular cluster. The first evidence comes from the extrapolation to globular clusters of the $M_{BH}-M_{Bulge}$ relation found for super massive black holes in galactic nuclei (for details see \citealt{magorrian1998}), which leads to the prediction of the existence of IMBHs.

The second hint is related to the discovery of the so called
ULXs, i.e. ultra-luminous, compact $X$-ray sources (with
luminosity greater than $\sim 10^{39}$ erg s$^{-1}$), which are
believed to be IMBHs rather than binaries containing a normal
stellar mass black hole (\citealt{miller2003}).


More indirect evidence of the existence of IMBHs comes from
the study of the central velocity dispersion of stars in specific
globular clusters. For example, by using the velocity dispersion
measurements, Gerssen et al. (\citeyear{gerssen2002}, \citeyear{gerssen2003}), and
Gebhardt et al. (\citeyear{gebhardt2002}) (but see also
\citealt{pooley2006}) proposed that IMBHs may exist in M15 and G1
(an M31 globular cluster) with masses about $10^3$-$10^4$
M$_{\odot}$ \footnote{
The observed brightness and the velocity
dispersion profiles of the G1 (\citealt{baum2003}) and M15 (\citealt{baum2005})
globular clusters can be
well fitted by usual evolutionary King models. When the
mass segregation effect is taken into account, a sharp increase in
the mass-to-light ratio towards the cluster core is found, thus
avoiding the necessity of a central IMBH.}

Objects of the IMBH size are predicted by detailed N-body simulations (see e.g. \citealt{potegiezwart2004}), according to which an IMBH forms as a consequence of merging of massive stars. Furthermore, at least for the cases of M 15 and 47 Tucane, the precise measurements of $P$ and $\dot{P}$ of four millisecond pulsars (with negative $\dot{P}$) has allowed \citet{depaolis} to put rather stringent upper limits to the mass of the central black hole of $\sim 10^3$ M$_{\odot}$.

In addition to the previous evidence, it is also expected that
globular clusters with a central IMBH are characterized by a cusp
in the inner stellar density profile (i.e. $\rho \propto
r^{-7/4}$), so that the projected density profile, as well as the
surface brightness, should also have a cusp with slope $-3/4$.
As shown by Miocchi (\citeyear{miocchi}), the globular
clusters that most likely harbor an IMBH are those having the
projected photometry well fitted by a King profile, except in the
central part where a power law deviation ($\alpha\simeq -0.2$)
from a flat behavior is expected. However, as pointed out by Baumgardt
et al. (\citeyear{baum2003}, \citeyear{baum2005}), a similar behavior is
also expected in a constant core density King profile ($\alpha\simeq 0$)
when the mass segregation effect of stellar remnants is considered.
This explanation, however, is uncertain because it depends
on the assumption that all the neutron stars and/or stellar
mass black holes are retained in the cluster so that the existence
of a central IMBH cannot be completely ruled out (see
\citealt{grh2005}). On the other hand, the errors with which the slopes of central densities
can be determined are of the order $0.1-0.2$ (\citealt{noyola}), so that surface density profiles do not give clear evidence of the existence of IMBH in globular clusters, thus requiring observations in different bands.

Among all the known globular clusters in our Galaxy, NGC 6388 is
one of the best candidates to host an IMBH  (\citealt{baumgardt2005}).
NGC 6388 is a globular cluster at distance $d\simeq 11.5$ kpc
(with center at coordinates A.R.$=17^{h} 36^{m} 17.6^{s}$,
DEC$=-44^{\degr} 44^{\arcmin} 08.2^{\arcsec}$) and with an
estimated mass of $2.6\times 10^6$ M$_{\odot}$
(\citealt{lanzoni2007}). By using a combination of high resolution
(HST ACS-HCR, ACS-WCS and WFPC2) and wild-field (ESO-WFI)
observations, Lanzoni et al. (\citeyear{lanzoni2007}) derived the
center of gravity, the projected density profile and the central
surface brightness profile of NGC 6388. While the overall
projected profile is well fitted by a King model (with
concentration parameter $c=1.8$ and core radius $r_c\simeq
7.2\arcsec$), a significant power law (with slope $\alpha\simeq
-0.2$) deviation from a flat core behavior has been detected
within $\simeq 1\arcsec$. This was interpreted as the signature of
the existence of an IMBH with mass $\simeq 5.7\times 10^3$
M$_{\odot}$ at the center of NGC 6388. We expect that the central
IMBH emits significant radiation in the $X$-ray band as a
consequence of the accretion of the surrounding matter. Thus, we
searched for both $XMM$-Newton and $Chandra$ observations towards
the globular cluster NGC 6388.

The paper is structured as follows: In Sect. 2, we briefly
describe the $XMM$-Newton and $Chandra$  observations and data
reduction. In Sect. 3, we describe the main characteristics of
the sources detected within a few core radii of the globular
cluster and in Sect. 4 we address our main results and
conclusions.

\section{{\it XMM}-Newton and $Chandra$ observation and data reduction}
The globular cluster NGC 6388 was observed in March 2003
(observation ID 0146420101) with both the EPIC MOS and PN cameras
(\citealt{mos,pn}) operating with the medium filter mode. The EPIC
observation data files (ODFs) were processed using the {\it
XMM}-Science Analysis System (SAS version $7.0.0$). Using the
latest calibration constituent files currently available, we
processed the raw data with the {\it emchain} and {\it epchain}
tools to generate calibrated event lists. After screening with
standard criteria, as recommended by the Science Operation Centre
technical note XMM-PS-TN-43 v3.0, we rejected any time period
affected by soft proton flares. The remaining time intervals
resulted in effective exposures of $\simeq 2.5\times10^4$ ks,
$\simeq 2.4\times10^4$ ks, and  $\simeq 1.3\times10^4$ ks for
MOS~1, MOS~2, and PN, respectively.

\begin{table*}
\begin{center}
\caption{We report some useful parameters for the globular cluster
NGC 6388 as given in \citet{harris}. The columns represent the
distance to NGC 6388, the core radius, the half mass radius, the
central luminosity density, the concentration as derived by a King
fit to the photometric data, the total magnitude and mass.}\label{table1}
\begin{tabular}{lllllll}
\hline
$D $(kpc)& $R_c$ (arcmin)& $R_H$ (arcmin)& $\log(\rho_c) {\rm (L_{\odot}/pc^3)}$ & $c$ & $V_t$ & $M({\rm \times10^6 M_{\odot}})$\\
\hline
   11.5      &  0.12     &  0.67     &    5.29                             & 0.17    & 6.72      & 2.6\\
\hline
\end{tabular}
\end{center}
\end{table*}
\begin{table*}
\begin{center}
\caption{Properties of the observed discrete sources in NGC 6388 (see text for the source 14*).}\label{table2}
\begin{tabular}{lclllll}
\hline
Source  & Net (counts)& HR1& HR2& HR3& $F_x^{\rm Abs}$ ($F_x^{\rm Cor}$) & $L_x^{\rm Abs}$ ($L_x^{\rm Cor}$)\\
        & (0.5-7 keV)            & ${\rm (S-M)/(S+M)}$   &  ${\rm(M-H)/(M+H)}$  &  ${\rm (S-H)/(S+H)}$  & $(\times 10^{-14}$ cgs) & $(\times 10^{32}$ cgs)\\
\hline
 1        &172.0~$\pm$~13.3&~0.26 $\pm$ 0.09&~0.20 $\pm$ 0.09&-0.06 $\pm$ 0.10&2.9~~(3.7)&4.6~~(5.8)               \\
2 ....... &195.1~$\pm$~14.2&-0.18 $\pm$ 0.11&-0.52 $\pm$ 0.07&-0.38 $\pm$ 0.07&3.3~~(4.2)&5.2~~(6.6)               \\
3 ....... &527.1~$\pm$~23.0&~0.53 $\pm$ 0.04&~0.96 $\pm$ 0.01&~0.86 $\pm$ 0.04&8.9~~(11.2)&14.0~~(17.6)            \\
4 ....... &~37.0~$\pm$~~6.3&-0.12 $\pm$ 0.20&-0.04 $\pm$ 0.21&~0.08 $\pm$ 0.20&0.6~~(0.8)&0.9~~(1.2)               \\
5 ....... &125.1~$\pm$~11.4&~0.80 $\pm$ 0.06&~0.95 $\pm$ 0.03&~0.60 $\pm$ 0.21&2.1~~(2.7)&3.3~~(4.2)               \\
6 ....... &~41.9~$\pm$~~6.8&-0.04 $\pm$ 0.19&-0.07 $\pm$ 0.20&-0.03 $\pm$ 0.20&0.7~~(0.9)&1.1~~(1.4)                \\
7 ....... &134.7~$\pm$~11.9&~0.66 $\pm$ 0.07&~0.89 $\pm$ 0.04&~0.57 $\pm$ 0.16&2.3~~(2.9)&3.6~~(4.6)               \\
8 ....... &~87.0~$\pm$~~9.8&~0.30 $\pm$ 0.13&~0.18 $\pm$ 0.13&-0.12 $\pm$ 0.16&1.4~~(1.8)&2.2~~(2.8)               \\
9 ....... &109.0~$\pm$~10.6&~0.25 $\pm$ 0.11&~0.27 $\pm$ 0.11&~0.02 $\pm$ 0.13&1.8~~(2.3)&2.8~~(3.6)               \\
10 ..... &355.0~$\pm$~19.1&~0.54 $\pm$ 0.05&~0.97 $\pm$ 0.02&~0.91 $\pm$ 0.03&6.0~~(7.6)&9.4~~(11.9)               \\
11 ..... &~78.1~$\pm$~~8.8&-0.23 $\pm$ 0.13&~0.02 $\pm$ 0.15&~0.25 $\pm$ 0.13&1.3~~(1.6)&2.0~~(2.5)                 \\
12 ..... &108.1~$\pm$~10.7&-0.18 $\pm$ 0.13&-0.11 $\pm$ 0.14&~0.07 $\pm$ 0.12&1.8~~(2.3)&2.8~~(3.6)                 \\
13 ..... &127.8~$\pm$~11.7&~0.14 $\pm$ 0.10&~0.30 $\pm$ 0.11&~0.17 $\pm$ 0.12&2.2~~(2.7)&3.4~~(4.2)                 \\
14* ... &789.6~$\pm$~28.8&~0.38 $\pm$ 0.04&~0.58 $\pm$ 0.03&~0.26 $\pm$ 0.06&13.4 (16.8)& 21.1~~(26.5)              \\
\hline
\end{tabular}
\end{center}
\end{table*}

\begin{figure}[htbp]
\vspace{7.0cm} \includegraphics{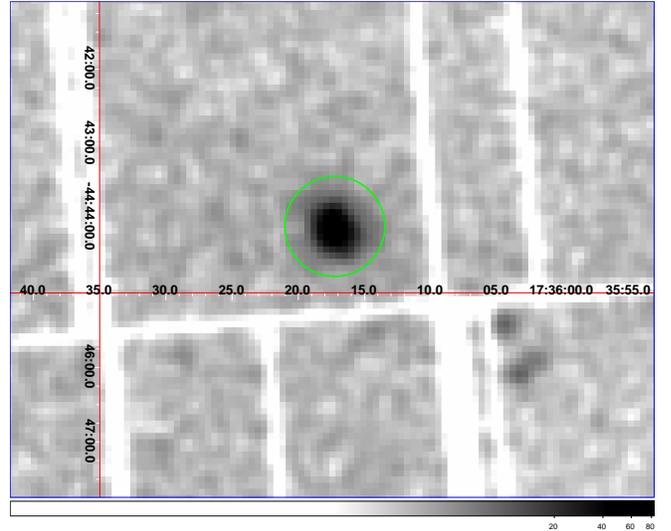}
\caption{The {\it XMM}-Newton field of view (PN camera) centered on NGC 6388. The green circle has a radius comparable to
the half mass radius ($\simeq 0.67\arcmin$) of the globular cluster.}
\label{fig_xmm}
\end{figure}

In Fig. \ref{fig_xmm}, the {\it XMM}-Newton observed field of view (PN camera) is shown. The green circle is centered on NGC 6388 and has a radius equal to the half mass radius of the globular cluster.
The source spectra were extracted in a circular region centered on the nominal position of the target in the
three EPIC cameras, while the background spectra were accumulated in annuli around the same coordinates. The resulting spectra were rebinned to have at least 25 counts per energy bin.

\begin{figure}[htbp]
\vspace{7.0cm} \includegraphics{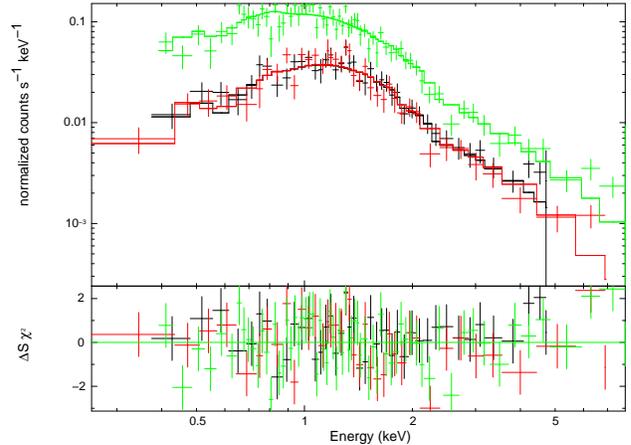}
\caption{The simultaneous fit to the MOS 1 (red), MOS 2 (black), and PN (green) data with a model
based on an absorbed power law (see text for more details).}
\label{fig_xmm_spectra}
\end{figure}

The spectra were simultaneously fitted with XSPEC (version 12.0.0). In Fig. \ref{fig_xmm_spectra},
we show the MOS 1, MOS 2, and PN spectra for NGC 6388 and the respective fits. The best-fitting model was an absorbed power law ($\chi^2/\nu$=1.14 for $\nu=147$). We left all the parameters free, yielding $N_H=2.7_{-0.3}^{+0.3}\times 10^{21}$ cm$^{-2}$, $\Gamma=2.4_{-0.1}^{+0.1}$ and $N=2.2_{-0.2}^{+0.2}\times 10^{-4}$ for the column density\footnote{Within the error, the column density obtained from the fit procedure is compatible with that due to gas in our Galaxy along the line-of-sight to NGC 6388, i.e. $N_H\simeq 2.5\times 10^{21}$ cm$^{-2}$ (\citealt{dickey}).}, power law index and normalization, respectively.

The flux in the 0.5-7 keV is $F_{0.5-7}=4.0_{-0.2}^{+0.2}\times10^{-13}$ erg cm$^{-2}$ s$^{-1}$ which, for the globular cluster distance of $\sim 11.5$ kpc, corresponds to a luminosity of $L_{0.5-7}\simeq 6.31\times10^{33}$ erg s$^{-1}$. Note that all the uncertainties quoted above are given at a $90\%$ confidence level.

The information that we can get from $XMM$-Newton data are important in obtaining
an overall description of the $X$-ray radiation coming from NGC 6388. However, the {\it Chandra} satellite has a much better angular resolution than the $XMM$-Newton telescope and therefore it is worth using {\it Chandra} to verify whether the emission detected by $XMM$-Newton is due to a single source rather than superpositions of several bright $X$-ray sources.

The globular cluster was observed  by {\it Chandra} with the ACIS-S camera (for $\simeq45$ ks, observation ID 5505). In our analysis, we used the event 2-type files and followed the standard procedures for analysis of {\it Chandra} data using the CIAO version 3.4 tool suite. The background level during the observation was nominal.

\begin{figure*}[htbp]
\vspace{0.2cm}
\begin{center}
$\begin{array}{c@{\hspace{0.1in}}c@{\hspace{0.1in}}c}
\epsfxsize=2.0in \epsfysize=2.0in \epsffile{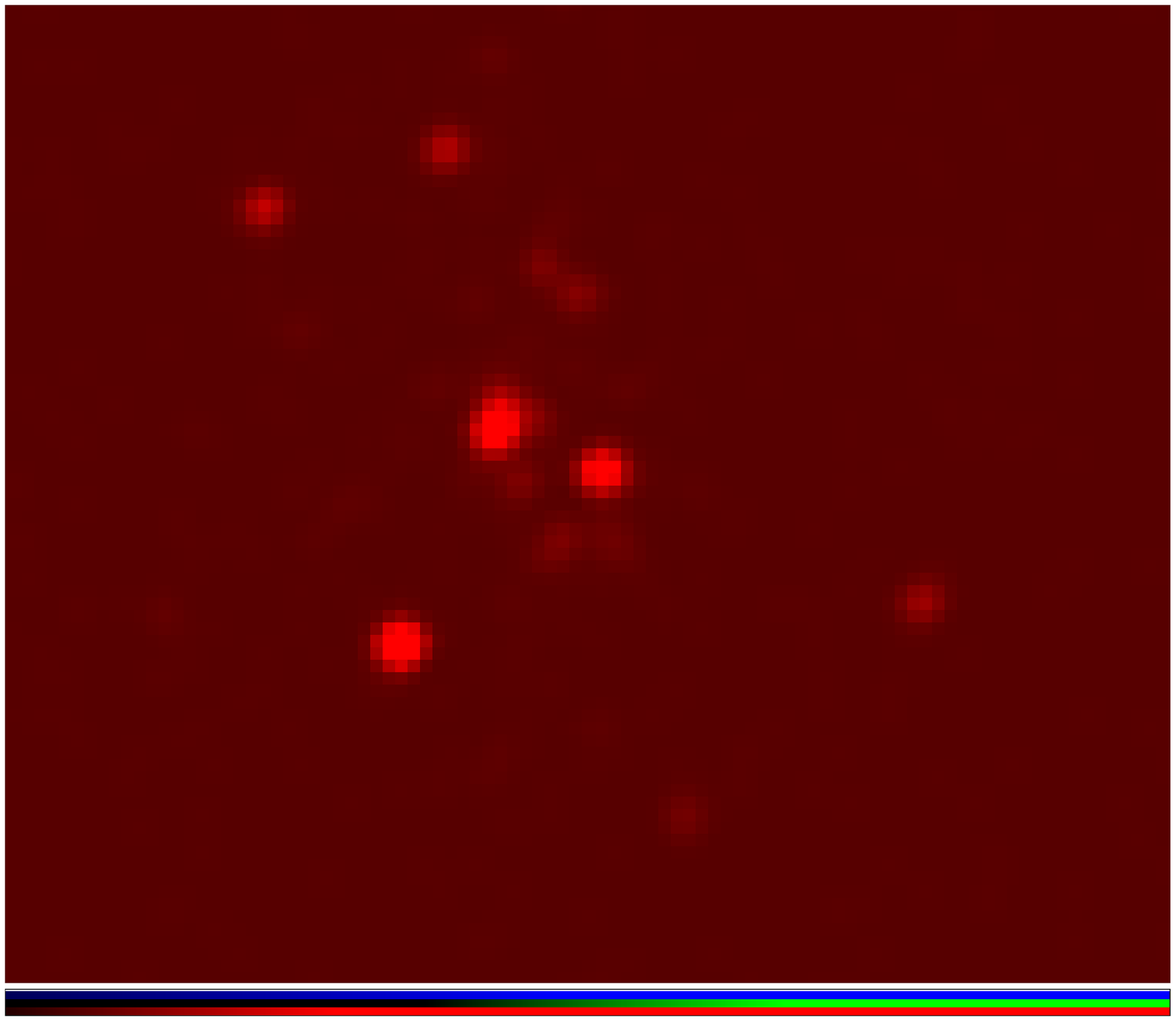} &
\epsfxsize=2.0in \epsfysize=2.0in \epsffile{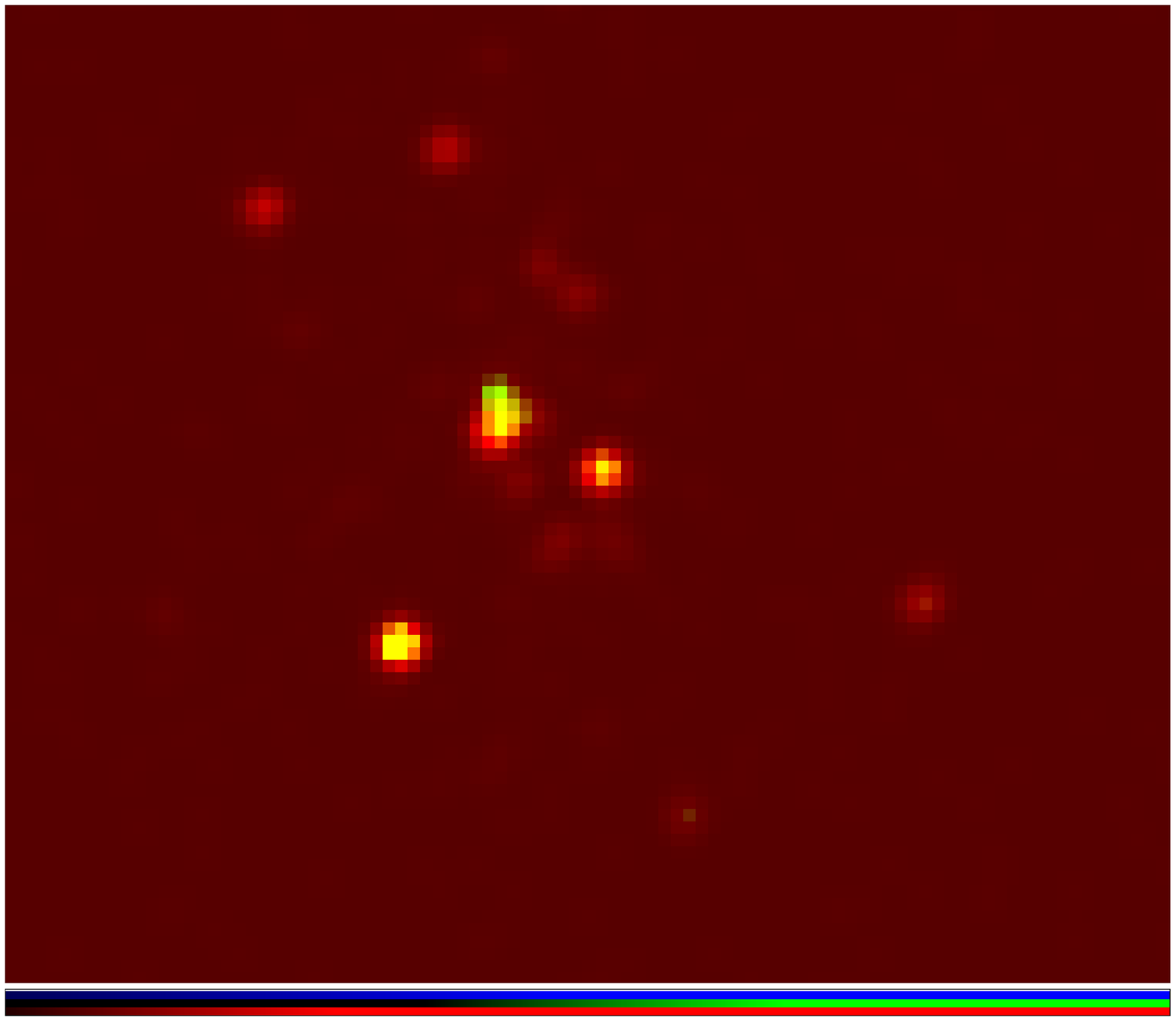} &
\epsfxsize=2.0in \epsfysize=2.0in \epsffile{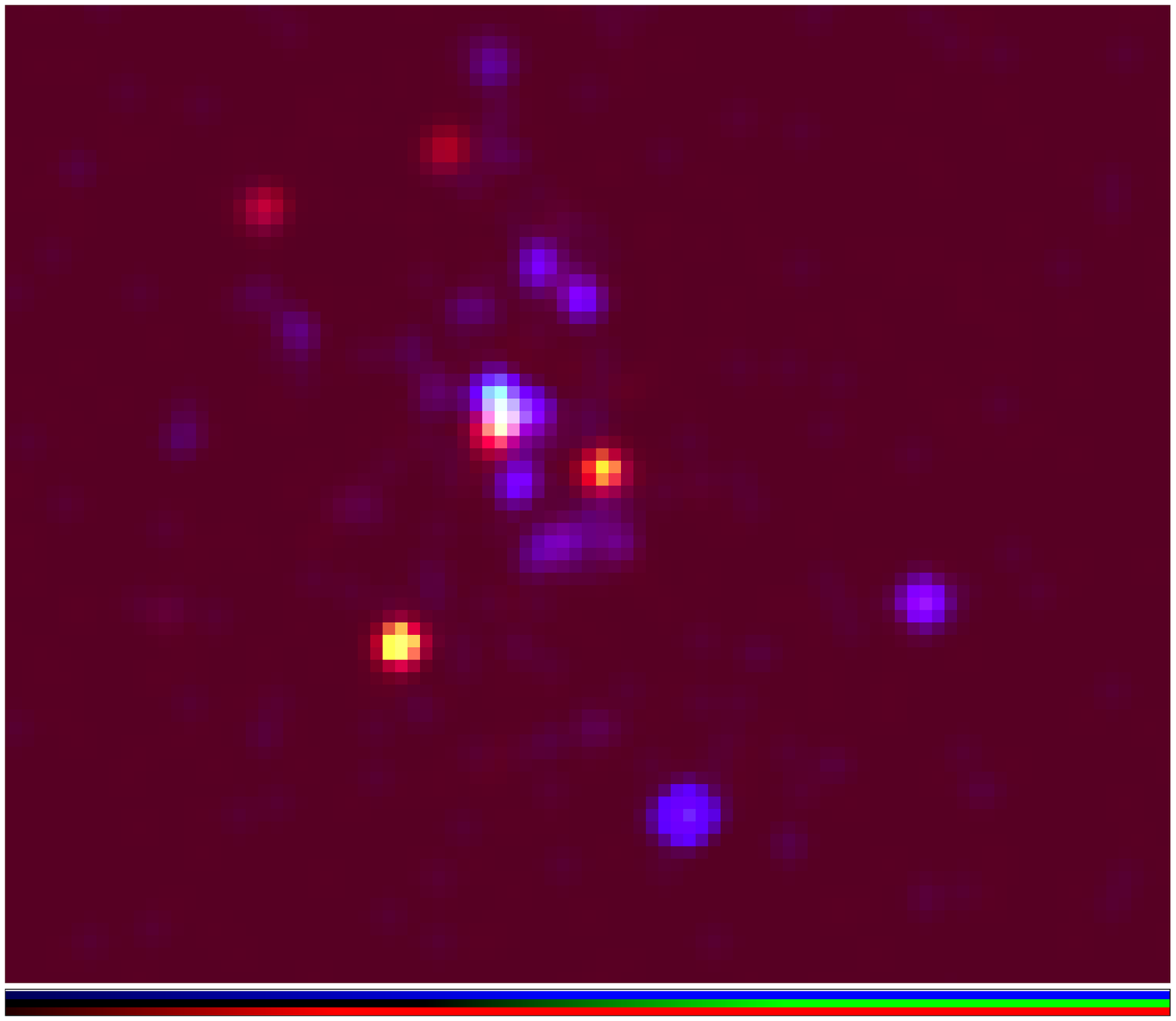} \\
\end{array}$
\end{center}
\caption{{\it Chandra}/ACIS images in the soft, medium and hard bands. From the left to the right, the soft, soft+medium and soft+medium+hard images are shown.}
\label{truecolor}
\end{figure*}

We created images in the full (F=0.5-7 keV), soft (S=0.5-1.5 keV), medium (M=1.5-2.5 keV) and hard (H=2.5-7 keV) bands, and created a true color $X$-ray image of the globular cluster (given in Fig. \ref{truecolor}). The $X$-ray emission towards NGC 6388 detected by the $XMM$-Newton satellite is associated with several discrete sources.

We searched in the images in each band for discrete sources by using the CIAO {\it celldetect} tool with a threshold signal-to-noise detection value of $3$. In the following, we will treat only the sources detected within $\sim 2$ core radii from the NGC 6388 center. This resulted in the detection of $16$ discrete sources well within the half mass radius ($\simeq 40\arcsec$) of NGC 6388. Of the detected sources, $9$ are contained within the globular cluster core radius ($\simeq 7\arcsec$),  while $3$ are close to the NGC 6388 center of gravity, which is located at A.R.$=17^{h} 36^{m} 17.23^{s}$ and DEC$=-44^{\degr} 44^{\arcmin} 07.1^{\arcsec}$ (\citealt{lanzoni2007}) with an uncertainty of $\simeq 0.3\arcsec$ in both coordinates. The absolute errors associated with {\it Chandra} astrometry is $\simeq 1\arcsec$.

The detected $X$-ray sources appear to be associated with the globular cluster
NGC 6388. Based on the Log N-Log S relationship of \cite{giacconi2001}, the estimated number of the background sources (with flux greater than the minimum detected flux, i.e. $\simeq 8\times10^{-15}$ erg cm$^{-2}$ s$^{-1}$) contained within the cluster half mass radius is $\simeq 10^{-2}$.

\begin{figure*}[htbp]
\vspace{8.0cm} \includegraphics{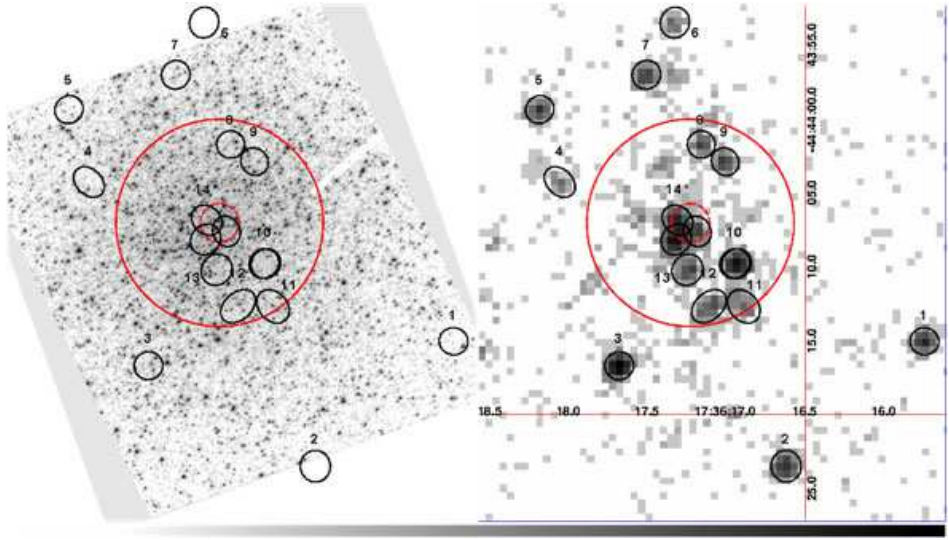}
\caption{A comparison of the {\it Chandra} and HST ACS-HRC fields of view towards NGC 6388 is given. The red solid circles represent the  globular cluster core with radius $\simeq 7.2\arcsec$, while the red dashed circles ($\simeq 1.3\arcsec$) are centered on the center of gravity (see text for more details). {\bf LOW RESOLUTION IMAGE (see the published version of the paper).}}
\label{ottico_chandra}
\end{figure*}

In Fig. \ref{ottico_chandra}, we show the detected discrete sources in the {\it Chandra} 0.5-7 keV image (right panel) as the encircled ones (each source being labeled with an increasing number). In left panel of the same figure, we show a $29.1\arcsec\times28.4\arcsec$ HST ACS-HRC image (filter V F555W) of the same field of view. The red solid circle represents the NGC 6388 core, while the dashed circle is centered on the globular cluster center of gravity and has a radius of $\simeq 1.3\arcsec$ as the sum of the uncertainties of the center of gravity and the absolute position error of {\it
Chandra}.

We estimated the counts of each source from an aperture including most of the observed emission\footnote{In the case of point-like sources, the aperture radius was $\simeq 1.3\arcsec$, which corresponds to an encircled energy of $\simeq 90\%$ at 1.4 keV.}. The background has been estimated by using annuli around each source when possible, or circles with the same extraction region radius otherwise (excluding any encompassed source).

For each source, we extracted the net number of counts in the full band and evaluated the hardness ratios ${\rm HR1=(S-M)/(S+M)}$, ${\rm HR2=(M-H)/(M+H)}$ and ${\rm HR3=(S-H)/(S+H)}$, where S, M and H correspond to the net counts in the soft, medium and hard bands, respectively. We give the results of the analysis in Table \ref{table2}, where
the counts in the 0.5-7 keV band, the hardness ratios and the absorbed ($F_x^{\rm Abs}$) and corrected ($F_x^{\rm Cor}$) fluxes are shown. In the last column, the absorbed ($L_x^{\rm Abs}$) and corrected luminosity ($L_x^{\rm Cor}$) of each source (in the 0.5-7 keV) has been determined assuming a $\Gamma=1.7$ power law absorbed by the Galactic line-of-sight column density $N_H\simeq 2.5\times 10^{21}$ cm$^{-2}$.

The three sources (labeled as 14*) present in the globular cluster center should be viewed cautiously: since we cannot resolve them in a better way we refer to the cumulative net counts and luminosity.

\section{The inner $X$-ray cluster in NGC 6388}

As we have seen in the previous Section, the better angular resolution of the {\it Chandra} satellite with respect to that of $XMM$-Newton, allowed us to detect several discrete sources within a few core radii of NGC 6388. The main properties of the detected sources are given in Table \ref{table2}.

Any classification of the detected sources requires an understanding of their spectral shapes. Due to the low count rates of most of the detected sources, formal spectral modeling is only possible for the brightest ones (namely those with more than  $300$ counts and labeled as 3, 10 and 14*).

\begin{figure}[htbp]
\vspace{6.5cm} \includegraphics{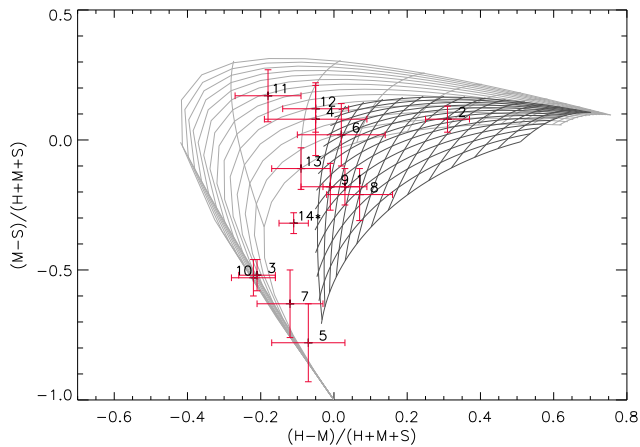}
\caption{Color-color diagram of the sources (data points) detected within a few core radii in NGC 6388 and color-color contours for bremsstrahlung (grey region) and power law (black region) components, respectively. On the horizontal and vertical axes, the two hardness ratios $HR_{\rm hard}=(H-M)/(S+M+H)$ and $HR_{\rm soft}=(M-S)/(S+M+H)$  are given (see text for details).}
\label{hardness}
\end{figure}

For the other sources, a rough classification can be done by using a color-color diagram (see Fig. \ref{hardness}) in which the two hardness ratios $HR_{\rm soft}=(M-S)/(S+M+H)$ and  $HR_{\rm hard}=(H-M)/(S+M+H)$ are given. In the same plot, we give the expected set of color-color contours for bremsstrahlung (grey region) and power law (black region) components, respectively.
In both the two regions, the equivalent hydrogen column $N_H$ (taken varying between $10^{20}$ cm$^{-2}$ and $10^{22}$ cm$^{-2}$) is associated with almost horizontally-oriented lines. The temperature $kT$ of the bremsstrahlung models (taken in the range 0.1-2.5 keV) is associated with primarily vertical-lines. In the case of the power law region, vertical orientation is associated with values of the parameter $\Gamma$ in the range 0.1-2.5.

According to the classification scheme of \citet{jenkins}, to
which we refer for more details, most of the sources with $HR_{\rm
soft}\ut>-0.2$ seem to be low mass $X$-ray binaries, with the
source 2 possibly being a high mass $X$-ray binary containing a
neutron star.

The source labeled as 14* is positioned in a region of the color-color diagram where its spectral properties are not well fitted by a single spectral component (represented by the two shaded regions), even if it seems marginally
consistent with power law models with large exponents ($\Gamma$ greater than 2.5, typical of $X$-ray emission from IMBHs, which are expected to be soft sources). This was expected since we extracted the cumulative spectrum of what appeared to be the superposition of three different sources (possibly LMXRBs), each of which has different spectrum properties.
Finally, note the existence of four soft sources corresponding to $HR_{\rm soft}\ut<-0.5$.

Interestingly, the cumulative luminosity of the sources detected in the {\it Chandra} field of view (see Table \ref{table2}) is fully consistent with that derived by the {\it XMM}-Newton observation of NGC 6388.
For the three sources with more than $300$ counts (3, 10 and 14*) detected within the NGC 6388 globular cluster, we extracted the spectrum from a circular aperture centered on each of the three sources. Note that the source labeled as 14* seems to be the superposition of three distinct sources (see also the true color images in Fig. \ref{truecolor}), for which it was not possible to extract the single spectra. In this case we refer to the cumulative spectrum.

\begin{figure*}[htbp]
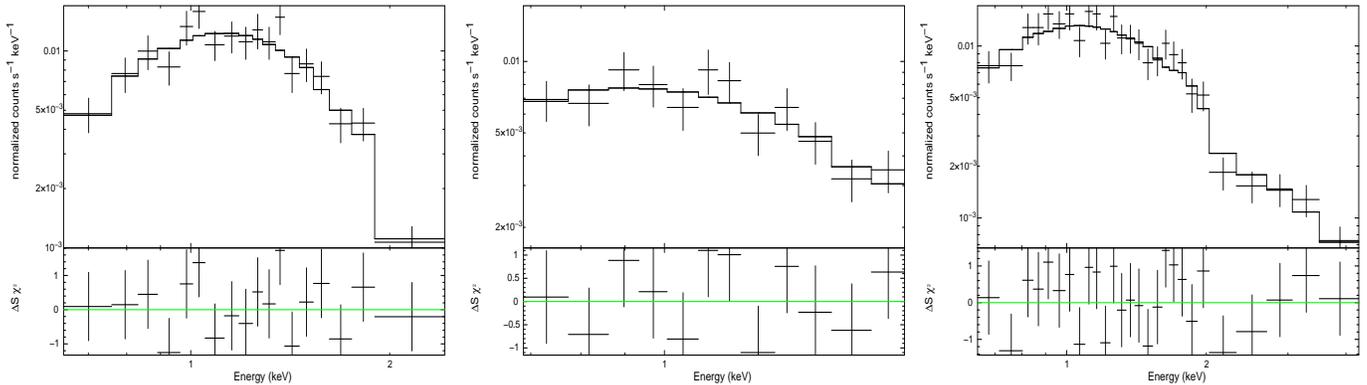

\vspace{0.2cm}
\begin{center}
\vspace{6.5cm}
$\begin{array}{c@{\hspace{0.1in}}c@{\hspace{0.1in}}c}
\includegraphics{8850f6a.ps} &
\includegraphics{8850f6b.ps} &
\includegraphics{8850f6c.ps} \\
\end{array}$
\end{center}
\caption{The spectra of the sources labeled as 3, 10 and 14* from the left to the right (see text for details on the fit). }
\label{spectra}
\end{figure*}

The spectra (shown in Fig. \ref{spectra}) were rebinned with 25 counts per bin and fitted using the XSPEC spectral fitting package. All the errors given below are at a $90\%$ confidence level.

The spectrum of the source 3 is well fitted ($\chi^2/\nu$=0. 76 for $\nu=15$) by an absorbed black-body model with temperature $kT=0.26_{-0.10}^{+0.03}$ keV, hydrogen column density $N_H=3.1_{-0.1}^{+0.1}\times 10^{21}$ cm$^{-2}$, and normalization $N=1.2_{-0.3}^{+0.6}\times 10^{-7}$ corresponding to a flux (in the 0.5-7 keV) of $4.5\times 10^{-14}$ erg cm$^{-2}$ s$^{-1}$.

Source 10 is characterized by a power law whose best fit parameters ($\chi^2/\nu$=0. 76 for $\nu=9$) are $N_H=1.6_{-1.6}^{+1.7}\times 10^{21}$ cm$^{-2}$, $\Gamma=2.4_{-1.0}^{+1.3}$, and $N=2.0_{-0.9}^{+0.3}\times 10^{-5}$ for the column density, power law index and normalization, respectively. The derived flux in the 0.5-7 keV is
$5.1_{-3.6}^{+0.2}\times 10^{-14}$ erg cm$^{-2}$ s$^{-1}$.

As discussed above, the source labeled as 14* seems to be the superposition of three different sources: one very soft and the other two harder, which lie close to the center of gravity of the globular cluster. With the present observations there is no way to discriminate whether one of these $X$-ray sources is associable with the IMBH hosted in NGC 6388.
Nevertheless, if this is the case, the observed flux in the 0.5-7 keV can be considered as the present upper limit to the IMBH $X$-ray signal. The source appears to be soft since the best fit absorbed power law ($\chi^2/\nu$=0.81 for $\nu=26$) has parameters $N_H=2.7_{-0.1}^{0.05}\times 10^{21}$ cm$^{-2}$, $\Gamma=2.4_{-0.2}^{+0.3}$ and $N=4.7_{-0.6}^{+1.4}\times 10^{-5}$ for the column density, power law index and normalization, respectively. In this case, the flux in the 0.5-7 keV would be $1.02\times 10^{-13}$ erg cm$^{-2}$ s$^{-1}$ corresponding to an $X$-ray luminosity of $2.7\times 10^{33}$ erg s$^{-1}$.

\section{Results and conclusions}

Intermediate-mass black holes may represent the link between the
stellar mass black holes present throughout the Milky Way and
the super-massive black hole thought to exist in the center of the
Galaxy and in external galaxies. Recent theoretical works have
suggested that all the globular clusters may have central black
holes with masses of the order $10^{-3}$ times the stellar mass in
the cluster, as a consequence of merging processes of stellar mass
black holes (\citealt{miller2002}). Present observational campaigns
seem to confirm this hypothesis.

In the particular case of NGC 6388, numerical simulations (see
e.g. \citealt{baumgardt2005}) have shown that it is a good
candidate to host an IMBH of a few $10^3$ M$_{\odot}$.
Interestingly, the optical HST observations of the globular
cluster and the detailed study of the brightness surface profile
down to a distance of $\simeq 1\arcsec$ from the center, revealed
a significant power law (with slope $\alpha\simeq -0.2$) deviation
from a flat core behavior (\citealt{lanzoni2007}). This was explained
assuming the existence of an IMBH with mass $5.7\times 10^3$
M$_{\odot}$ in the center of the globular cluster.

As a consequence of matter accretion, we expect that the putative
IMBH should accrete and emit radiation in the $X$-ray band, so
we searched for such a signature in the observations
available in both $XMM$-Newton and {\it Chandra} archives.

The study of the central region of NGC 6388 in the 0.5-7 keV
energy band reveals the existence of several discrete $X$-ray
sources (see Sect. 3 for a detailed discussion) among which
three of them seem to be overlapped and close to the center of
gravity of NGC 6388. Although we can not obtain the spectrum of
each of the three sources separately, we have speculated that if one of these
is the putative NGC 6388 IMBH, the observed
$X$-ray signal can be thought as an upper limit to the IMBH flux.

Hence, with reference to Table \ref{table2}, the unabsorbed
$X$-ray flux (in the 0.5-7 keV band) of the IMBH is
$F_X^{Obs}\ut<1.6\times 10^{-13}$ erg cm$^{-2}$ s$^{-1}$,
corresponding to a luminosity of $L_X^{Obs}\ut<2.7\times 10^{33}$
erg s$^{-1}$ for the NGC 6388 distance of 11.5 kpc.

At this point one can evaluate the IMBH radiative efficiency
$\eta=L_X/L_{Edd}$ with respect to the maximum allowed black hole
accretion rate given by the Eddington luminosity $L_{Edd}\simeq
1.38\times10^{38}(M/M_{\odot})$ erg s$^{-1}$. For the IMBH at the
center of NGC 6388  one gets $L_{Edd}\simeq7.87\times 10^{41}$ erg
s$^{-1}$, so that we can conclude that it is accreting  with
efficiency $\eta \ut< 3\times10^{-9}$. Note that this accretion
efficiency is in agreement with the efficiency estimates for black
hole accretion in quiescent galaxies and ultra-low luminous AGNs,
for which $\eta$ is typically in the range $ 4\times
10^{-12}-6\times 10^{-7}$ (\citealt{baganoff}).

The bolometric luminosity of the NGC 6388 IMBH can be inferred
from the broadband spectral energy distributions of galactic
nuclei (see Elvis et al. \citeyear{elvis} for details). In this
case, it is found that the bolometric correction for the $X$-ray
band corresponds to a factor $\sim 7-20$, so that
$L_{bol}/L_{Edd}\ut<(2-6)\times 10^{-8}$ (or  $L_{bol} \ut<
(2-5)\times 10^{34}$ erg s$^{-1}$).

It could be also interesting to estimate the expected accretion
luminosity ($L_{acc}$) of the IMBH as a consequence of the
accretion of the surrounding gas, and compare it to the bolometric
luminosity ($L_{bol}$) above. It is indeed expected that
post-main-sequence stars continuously lose mass that is injected
both in the cluster and intracluster medium. Dispersion
measurements derived from radio observations of pulsars give the
most sensitive probe of the gas content in globular clusters.
Studying the population of millisecond pulsars in the globular
clusters M 15 and 47 Tuc, Freire et al. (\citeyear{freire}) find
indications of the presence of a plasma with electron density
$n_e\simeq 0.2$ atoms cm$^{-3}$ and temperature $T\simeq 10^4$ K (see
also Ho at al. \citeyear{ho} for a detailed study
of the M 15 globular cluster). In common with other authors (\citealt{maccarone}),
we assume in the following that the gas density in NGC 6388
is $\sim 0.2$ atoms cm$^{-3}$. If the IMBH at the center of NGC 6388
accretes spherically through the Bondi accretion process, the
gravitational potential of the IMBH dominates the dynamics of the
gas within the accretion radius defined as $R_a\simeq
GM_{bh}/c_s^2$, where $c_s\simeq 0.1T^{1/2}$ km s$^{-1}$ is the
sound speed in a plasma at temperature $T$. For $T=10^4$ K and
$M_{bh}=5.7\times 10^{3}$ M$_{\odot}$, one obtains $c_s\simeq 10$
km s$^{-1}$ and $R_a\simeq 0.3$ pc. The accretion mass rate in the
Bondi mechanism is $\dot{M_{B}}=4\pi R_a^2 \rho _a c_s$, where
$\rho _a$ is the gas density at $R_a$. Assuming that the efficiency
in converting the accreted mass into energy is the standard
$\epsilon=10\%$, the expected luminosity due to accretion is
$L_{acc}=\epsilon \dot{M_B}c^2$, i.e.
\begin{equation}
\begin{array}{l}
L_{acc}=\displaystyle{2.4\times 10^{38} \left(\frac{\epsilon}{0.1}\right)}\times \\ \\
\displaystyle{\times\left(\frac{M_{bh}}{5.7\times 10^3~{\rm M_{\odot}}}\right)^2
\left(\frac{n}{0.2~{\rm cm^{-3}}}\right)\left(\frac{c_s}{10~{\rm km s^{-1}}}\right)^{-3}}~{\rm cgs}.
\end{array}
\end{equation}
Hence, in the case of the NGC 6388 IMBH, the accretion luminosity
is clearly larger than the bolometric luminosity quoted above by a
factor of at least $\simeq 10^4$, leading us to conclude that the
accretion efficiency is $\epsilon \ut < 8\times 10^{-5}- 2\times 10^{-4}$.

In recent years it has also been proposed that a
relationship between black hole mass, $X$-ray luminosity and radio
luminosity does exist (see e.g. \citealt{merloni} and
\citealt{maccarone}). In this context, IMBHs at the center of many
globular clusters, such as NGC 6388, may be easily identifiable objects
in deep radio observations.

\begin{figure}[htbp]
\vspace{6.5cm} \includegraphics{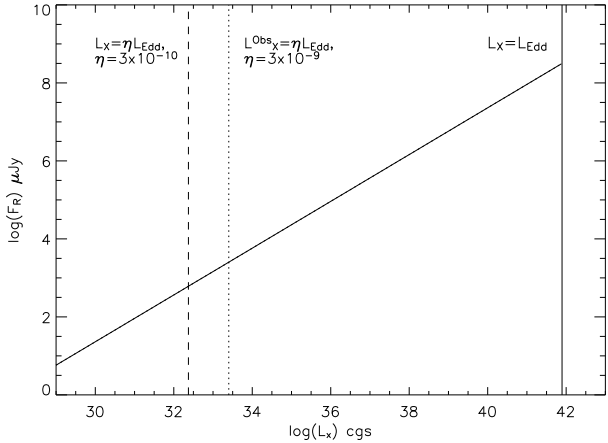} \caption{The expected radio
emission (solid oblique line) from the IMBH at the center  of NGC
6388 as a function of the IMBH $X$-ray luminosity. Present $X$-ray
observations can put only an upper limit of $\simeq 3$ mJy to the
black hole radio luminosity.} \label{radiofig}
\end{figure}

In particular, the expected radio flux at $5$ GHz from the putative IMBH in NGC 6388 would be
\begin{equation}
F_{5}=10\left(\frac{L_X}{3\times10^{31}{\rm cgs}}\right)^{0.6}\left(\frac{M_{bh}}{100{\rm M_{\odot}}}\right)^{0.78}\left(\frac{10{\rm kpc}}{d}\right)^{2}{\rm \mu Jy}.
\label{radio}
\end{equation}
Assuming that it is a quiescent and stable accreting black hole,
we can be more predictive about its radio luminosity. In Fig.
\ref{radiofig}, the expected radio emission (solid oblique line) from
the IMBH at the center of NGC 6388 is shown. The solid vertical
line represents the maximum allowed $X$-ray luminosity for an
Eddington limited accreting black hole. As one can see, depending
on the accretion efficiency (dotted and dashed lines are for $\eta
= 3\times10^{-9}$ and $\eta = 3\times10^{-10}$, respectively)  the radio flux at
5 GHz is $\ut< 3$ mJy, which is within the detection possibilities
of the Australia Telescope Compact Array (ATCA). Deep radio
observations within the core radius of NGC 6388 would also be
important for the possibility of discovering millisecond pulsars
nearby the cluster center that may allow a further and
independent constraint on the IMBH mass and position with respect
to the cluster center \footnote{Indeed, one expects that an IMBH
randomly moves, within the globular cluster core, due to the interaction with the other stars (assumed
to have the same mass $m$) with an amplitude $\sim r_c(m/M_{bh})$
(see e.g. \citealt{bw}, \citealt{gurzadyan} and
\citealt{merritt}).}.

\begin{acknowledgements}
This paper is based on observations from $XMM$-Newton, an
ESA science mission with instruments and contributions directly funded by ESA
member states and NASA. We are grateful to G. Trincheri and L. Bello for useful discussions. We are grateful to the
anonymous referee for the suggestions that improved the manuscript.
\end{acknowledgements}


\end{document}